\documentclass[conference]{IEEEtran}
\IEEEoverridecommandlockouts
\usepackage{cite}
\usepackage{xcolor}
\usepackage{amsmath}
\usepackage{amssymb}
\usepackage{amsthm}
\usepackage{amsfonts}
\usepackage{algorithm}
\usepackage{import}
\usepackage{braket}
\usepackage{qcircuit}
\usepackage{algpseudocode}
\usepackage{graphicx}
\usepackage{enumitem}
\usepackage{hyperref}
\hypersetup{hidelinks}
\usepackage{booktabs}
\usepackage{balance}
\usepackage{url}
\usepackage[normalem]{ulem}

\def\BibTeX{{\rm B\kern-.05em{\sc i\kern-.025em b}\kern-.08em
    T\kern-.1667em\lower.7ex\hbox{E}\kern-.125emX}}
\begin{document}

\title{Distributed Variational Quantum Linear Solver}

\author{
\IEEEauthorblockN{
Chao Lu\IEEEauthorrefmark{1},\quad
Pooja Rao\IEEEauthorrefmark{2},\quad
Muralikrishnan Gopalakrishnan Meena\IEEEauthorrefmark{1},\quad
Kalyana Chakaravarthi Gottiparthi\IEEEauthorrefmark{1}
}
\IEEEauthorblockA{
\IEEEauthorrefmark{1}\textit{National Center for Computational Sciences},
\textit{Oak Ridge National Laboratory}, Oak Ridge, TN, USA\\
\IEEEauthorrefmark{2}\textit{NVIDIA Corporation}, Santa Clara, CA, USA\\
Email: \{luc1, gopalakrishm, gottiparthik\}@ornl.gov;\ porao@nvidia.com
}
}


\maketitle

\begin{abstract}
The Variational Quantum Linear Solver (VQLS), a hybrid quantum-classical algorithm for solving linear systems, faces a practical scalability bottleneck: the Linear Combination of Unitaries (LCU) decomposition requires $\mathcal{O}(L^2)$ circuit evaluations per optimizer iteration, where $L$ can grow to $4^n$ in the worst case for an $n$-qubit system.
We address this computational bottleneck through two complementary strategies. First, we present a distributed VQLS (D-VQLS) framework\footnote{https://code.ornl.gov/olcf-qcfd/DVQLS.git}, built on NVIDIA CUDA-Q, that enables asynchronous, scalable distribution of the $\mathcal{O}(L^2)$ cost evaluations. Second, a fast Walsh--Hadamard transform (FWHT)-based Pauli decomposition with coefficient-amplitude pruning threshold $\tau=0.01$ curbs LCU growth for the structured Toeplitz family, reducing $L$ from $\mathcal{O}(2^n)$ to 64 for $n>6$ and compressing the circuit complexity per optimizer iteration from $\mathcal{O}(n4^n)$ to $\mathcal{O}(n)$. We derive the exact top-$L$ Frobenius error and connect it to worst-case solution error. For a 10-qubit tridiagonal Toeplitz system, the $L=64$ pruning yields a $256\times$ reduction---from 23 million to 90k circuits per optimizer iteration.

The D-VQLS framework is validated on the NERSC Perlmutter supercomputer using multi-node, multi-GPU ideal state-vector simulations, achieving over $99.99\%$ fidelity against classical solutions on tridiagonal Toeplitz and Hele--Shaw flow benchmarks, with near-ideal strong scaling up to 24 GPUs and $95.3\%$ weak scaling efficiency at 96 GPUs processing more than 360k circuits per optimizer iteration (from larger-$L$ pruning) for the 10-qubit system. Systematic profiling identifies the optimal resource allocation for distributed quantum circuit workloads, yielding a $2.52\times$ speedup for the configurations studied.
\footnote{\textbf{Notice:} This manuscript has been authored by UT-Battelle, LLC, under contract DE-AC05-00OR22725 with the US Department of Energy (DOE). The US government retains and the publisher, by accepting the article for publication, acknowledges that the US government retains a nonexclusive, paid-up, irrevocable, worldwide license to publish or reproduce the published form of this manuscript, or allow others to do so, for US government purposes. DOE will provide public access to these results of federally sponsored research in accordance with the DOE Public Access Plan (\href{http://energy.gov/downloads/doe-public-access-plan}{http://energy.gov/downloads/doe-public-access-plan}).}

\end{abstract}


\section{Introduction}

Solving large-scale linear systems of the form $A\mathbf{x}=\mathbf{b}$ remains a central computational bottleneck across scientific and engineering disciplines, from computational fluid dynamics (CFD) and structural analysis to weather forecasting, computational biology, and financial modeling. Classical iterative solvers exploit matrix sparsity and structure through targeted matrix-vector operations and preconditioning, yet their complexity scales polynomially with respect to the system size $N$ as $\mathcal{O}(N^m)$ with $m \in[1,2.3]$ depending on the algorithmic regime and matrix structure~\cite{coppersmith1982asymptotic,trefethen2022numerical}. As simulation fidelity demands grow and mesh resolutions increase, even the state-of-the-art classical linear solvers face fundamental scalability limitations, particularly when condition numbers grow rapidly with refinement or when many right-hand sides must be solved repeatedly across time steps.

Quantum computing offers a fundamentally different paradigm for tackling these systems. The Harrow--Hassidim--Lloyd (HHL) algorithm~\cite{harrow2009quantum} was the first to demonstrate a potential exponential speedup, achieving logarithmic scaling $\mathcal{O}(\log N)$ with respect to the matrix dimension. However, the HHL algorithm's reliance on deep and high-precision Quantum Phase Estimation (QPE) renders it impractical for the current ``Noisy Intermediate-Scale Quantum'' (NISQ) era~\cite{preskill2018quantumNISQ}. Several recent approaches have sought to bridge this gap, including adiabatic-inspired solvers~\cite{subacsi2019quantum, an2022quantum, costa2022optimal}, eigenstate filtering techniques~\cite{lin2020optimal}, parallel quantum-circuit generation~\cite{lu2025lugo}, and the Quantum Singular Value Transformation (QSVT) framework~\cite{gilyen2019quantum, chakraborty2018power}. While QSVT provides optimal asymptotic complexity, its circuit depth and block-encoding requirements~\cite{camps2024explicit, sunderhauf2024block} remain challenging for near-term quantum computers.

In response, the Variational Quantum Linear Solver (VQLS) has emerged as a promising hybrid quantum-classical alternative~\cite{bravo2023variational}. The approach reformulates the linear system as a variational optimization problem that minimizes a cost function $C(\theta)$ quantifying the distance between the prepared state $A\ket{x(\theta)}$ and the target state $\ket{b}$. VQLS uses shallow parameterized circuits suitable for current NISQ devices. The original work~\cite{bravo2023variational} demonstrated VQLS through numerical simulations of structured matrices as large as $2^{50} \times 2^{50}$ and hardware execution on Rigetti devices for $2^{10} \times 2^{10}$ systems. Subsequent studies have applied VQLS to advection--diffusion equations~\cite{demirdjian2022variational}, hydrological systems~\cite{golden2022quantum}, and incompressible fluid flows~\cite{ye2024hybrid, bosco2024demonstration, meena2025assessing}.

However, these demonstrations have largely been confined to matrices with analytically known unitary decompositions, which limits their applicability to real-world problems with arbitrary coefficient matrices. Encoding a general matrix requires a Linear Combination of Unitaries (LCU) decomposition, $A = \sum_{l=1}^L c_l A_l$, where each $A_l$ is unitary. In the worst-case scenario, a complete Pauli decomposition requires $4^n$ Pauli terms. This decomposition introduces a severe computational bottleneck: evaluating the VQLS cost function requires $\mathcal{O}(L^2)$ independent expectation values per optimizer iteration, each demanding a separate quantum-circuit execution. In our two-dimensional (2D) Hele--Shaw flow CFD benchmark, the number of LCU terms grows rapidly with problem size, producing tens of thousands of circuits per optimizer iteration and millions of circuit executions across a full optimization trajectory even for a small grid.

To address the increasing LCU-term count, we present the \textit{Distributed VQLS} (D-VQLS) framework, which exploits the embarrassingly parallel structure of LCU-based cost evaluation. D-VQLS combines NVIDIA CUDA-Q~\cite{kim2023cuda}, an open-source platform for hybrid quantum-classical computing, with MPI-based multi-node distribution. We demonstrate the framework on the National Energy Research Scientific Computing Center (NERSC) Perlmutter supercomputer. By distributing the $\mathcal{O}(L^2)$ circuit evaluations across multiple GPUs and nodes, we achieve near-ideal strong scaling up to 24 GPUs for a $2^{10}\times2^{10}$ problem (11-qubit circuits) and 95.3\% weak-scaling efficiency at 96 GPUs with 360,448 circuits per optimizer iteration. We further use NVIDIA Nsight Systems profiling to identify the optimal MPI task configuration. We want to emphasize that this framework could be applied to a distributed quantum processing unit (QPU) system when it is available in the future.


The major contributions of this work are summarized as follows:

\begin{itemize}
    \item To our knowledge, we present the first distributed framework for the LCU-based VQLS cost function, demonstrated at HPC scale. Built on NVIDIA CUDA-Q, the framework is demonstrated on multi-node GPU systems and is applicable to any backend supporting parallel circuit execution, including future distributed QPU architectures.
    \item We apply an output-efficient $\mathcal{O}(N^2\log N)$ FWHT-based Pauli decomposition with coefficient-amplitude pruning threshold $\tau=0.01$. For the 10-qubit tridiagonal Toeplitz system, this retains 64 terms and compresses the circuit count per optimizer iteration from 23 million to 90,112, leading to a $256\times$ reduction. We derive an exact top-$L$ Frobenius criterion and report the accuracy--cost tradeoff across $L \in [4, 128]$.
    \item We provide raw circuit-resource estimates for the tridiagonal Toeplitz benchmark up to $2^{20}\times2^{20}$, reporting pruned LCU terms, per-circuit gate counts, and circuit evaluations per optimizer iteration. These estimates quantify algorithmic circuit requirements before error correction and should not be interpreted as fault-tolerant resource estimates.
    \item The D-VQLS framework is validated on both a tridiagonal Toeplitz system and a Hele--Shaw flow benchmark, achieving over $99.99\%$ fidelity against classical solutions. For the 10-qubit tridiagonal Toeplitz system, we demonstrate near-ideal strong scaling up to 24 GPUs and $95.3\%$ weak scaling efficiency at 96 GPUs on NERSC Perlmutter, establishing that VQLS with LCU-encoded matrices scales efficiently on current supercomputing resources.
    \item Through detailed NVIDIA Nsight Systems profiling, we characterize the optimal resource allocation on HPC systems for our problem instances, identifying four CUDA MPS clients per A100 GPU as the optimal configuration on NERSC Perlmutter and achieving a $2.52\times$ speedup over the single-client-per-GPU baseline.
\end{itemize}

The remainder of this paper is organized as follows. Section~\ref{sec:methodology} presents the VQLS algorithm, the LCU decomposition strategy, the distributed execution framework, and the two benchmark applications used in this study. Section~\ref{sec:result} reports the validation, scaling, and profiling results. Section~\ref{sec:conclusion} discusses the implications of these findings and outlines directions for future work.

\section{Methodology} \label{sec:methodology}

This section presents the proposed D-VQLS framework in details. We begin with the VQLS formulation and its cost function, describe the LCU decomposition strategy for encoding arbitrary matrices, detail the distributed asynchronous execution framework built on CUDA-Q and MPI, and conclude with the two benchmark applications used for validation and performance analysis.

\subsection{Variational Quantum Linear Solver (VQLS)}

The Variational Quantum Linear Solver (VQLS) is a hybrid quantum-classical algorithm designed to solve the linear system $A\vec{x} = \vec{b}$ by mapping the problem onto a quantum Hilbert space~\cite{bravo2023variational}. The algorithm seeks to prepare a quantum state $\ket{x(\theta)} = V(\theta)\ket{0}$ such that $A\ket{x(\theta)} \propto \ket{b}$, where $A$ is a $2^n \times 2^n$ matrix, $\ket{b}$ is a normalized input state prepared by a fixed unitary $U_b\ket{0} = \ket{b}$, and $V(\theta)$ is a quantum ansatz (a quantum circuit with gate operations parameterized by $\theta$).

To guide the optimization of the parameters $\theta$, a cost function $C(\theta)$ quantifies the distance between the prepared state $A\ket{x(\theta)}$ and the target state $\ket{b}$, defined as:
\begin{equation}
C(\theta) = 1 - \frac{|\bra{b}A\ket{x(\theta)}|^2}{\bra{x(\theta)}A^\dagger A\ket{x(\theta)}}
\end{equation}

To improve convergence and mitigate barren plateau issues in deep circuits, a local cost function is employed~\cite{bravo2023variational}:
\begin{equation} \label{equ:local_cost}
    C_L(\theta)
=
1 - \frac{1}{n}
\sum_{j=1}^{n}
\frac{
\langle x(\theta) \vert
A^\dagger U_b P_j U_b^\dagger A
\vert x(\theta) \rangle
}{
\langle x(\theta) \vert A^\dagger A \vert x(\theta) \rangle
}
\end{equation}
where $U_b$ denotes the state-preparation unitary satisfying $U_b\ket{0} = \ket{b}$, and $P_j = \ket{0}\bra{0}_j \otimes I_{\bar{j}} = \tfrac{1}{2}\left(I + Z_j\right)$ is the projector onto $\ket{0}$ on the $j$-th qubit, with $Z_j$ the Pauli-$Z$ operator on that qubit and $I_{\bar{j}}$ the identity on the remaining $n-1$ qubits. This formulation guarantees $C_L(\theta) = 0$ if and only if $\ket{x(\theta)}$ is the exact solution up to a global normalization constant. In practice, the terms $\bra{b}A\ket{x(\theta)}$ and $\bra{x(\theta)}A^\dagger A\ket{x(\theta)}$ are evaluated via Hadamard tests, requiring only single-qubit measurements.

\subsection{Linear Combination of Unitaries (LCU) Decomposition}

For a matrix $A$ without an analytically known unitary decomposition, evaluating the VQLS cost function requires decomposing $A$ into a Linear Combination of Unitaries (LCU):
\begin{equation}
A = \sum_{l=1}^{L} c_l A_l
\end{equation}
where $A_l$ are unitary operators and $c_l$ are the corresponding complex coefficients. In the general case, the decomposed unitaries are Pauli strings drawn from the set $\{I, X, Y, Z\}^{\otimes n}$. The choice of decomposition strategy critically impacts the algorithm's efficiency, as the number of terms $L$ directly governs the number of quantum circuits required for each optimizer iteration.

To avoid the bottlenecks in the algorithmic complexity and memory requirements of Pauli decomposition, recent literature has explored domain-specific LCU strategies. Loaiza et al.~\cite{loaiza2025majorana} introduced low-rank tensor factorizations for electronic structure problems in quantum chemistry, reducing the required number of unitaries at the cost of expensive classical preprocessing and deep state-preparation circuits. For differential equations, Hogancamp et al.~\cite{hogancamp2026linear} constructed LCU decompositions for discrete elliptic operators with linear scaling in the spatial dimension, independent of the discretization grid size, though this approach is constrained to specific operator classes.

In this work, we adopt the fast Walsh--Hadamard transform (FWHT)-based Pauli decomposition to decompose an arbitrary matrix of dimension $N=2^n$~\cite{georges2025paulidecomp}. For an $N\times N$ matrix, the Pauli basis contains $N^2=4^n$ possible strings, but $4^n$ nonzero terms are required only in the worst case; structured matrices can have substantially shorter decompositions. A complete FWHT evaluates this basis in $\mathcal{O}(N^2\log N)=\mathcal{O}(n4^n)$ time and $\mathcal{O}(N^2)$ storage for the worst-case coefficient output and avoids a dense matrix factorization. The tridiagonal Toeplitz family studied here has only $2^n$ nonzero terms before pruning~\cite{anthropicClaudeOpus48,openaiCodex2026}.

Substituting the LCU decomposition into the local cost function from Equation~\eqref{equ:local_cost} yields the fully expanded form:
\begin{equation} \label{equ:expanded_cost}
C_L = 1 - \frac{1}{n} \sum_{j=1}^n \frac{\sum_{l=1}^L \sum_{l'=1}^L c_l^* c_{l'} \langle x(\theta) | A_l^\dagger U_b P_j U_b^\dagger A_{l'} | x(\theta) \rangle}{\sum_{l=1}^L \sum_{l'=1}^L c_l^* c_{l'} \langle x(\theta) | A_l^\dagger A_{l'} | x(\theta) \rangle}.
\end{equation}

This expression requires $n \cdot L^2$ terms to compute the numerator and $L^2$ terms to compute the denominator, leading to $(n+1) \cdot L^2$ distinct expectation values per optimizer iteration. Each expectation value is evaluated via a Hadamard test, which requires separate circuits for the real and imaginary components, yielding a total of $2(n+1) \cdot L^2$ circuit submissions per optimizer iteration. The circuit count per optimizer iteration therefore depends on both the number of LCU terms $L$ and the number of qubits $n$, establishing the computational scaling of the algorithm.

The decomposition is exact for an arbitrary input matrix before pruning, and D-VQLS is matrix-agnostic once the resulting Pauli strings are available. Compressibility, however, is governed by the Pauli-coefficient spectrum rather than by matrix size, sparsity, or condition number alone. A sparse matrix with isolated entries can have a flat Pauli spectrum, while a dense matrix can be concentrated on a few strings. Consequently, $4^n$ terms remain the worst-case upper bound.

To make the pruning criterion explicit, let $\{\sigma_l\}_{l=1}^{4^n}$ be the Pauli basis from which the unitaries $A_l$ are drawn, with
\begin{equation}
c_l=\frac{1}{N}\operatorname{Tr}(\sigma_l^\dagger A),
\qquad
\operatorname{Tr}(\sigma_l^\dagger \sigma_{l'})=N\delta_{ll'},
\end{equation}
where $c_l$ is the coefficient of Pauli string $\sigma_l$. We use absolute coefficient-magnitude pruning: starting from $L_{\rm pre}$ nonzero terms, we retain every term with $|c_l|>\tau$, leaving $L_{\rm post}$ terms. This preprocessing requires one pass over the $M\le4^n$ nonzero coefficients, or $\mathcal{O}(M)$ time. The generic complete-FWHT preprocessing has worst-case cost $\mathcal{O}(n4^n)$ because it evaluates the full Pauli basis. The retained-term count is determined by the coefficient spectrum; matrix dimension, sparsity, and condition number alone do not determine it.

\begin{figure*}[t]
    \centering
    \includegraphics[width=\textwidth]{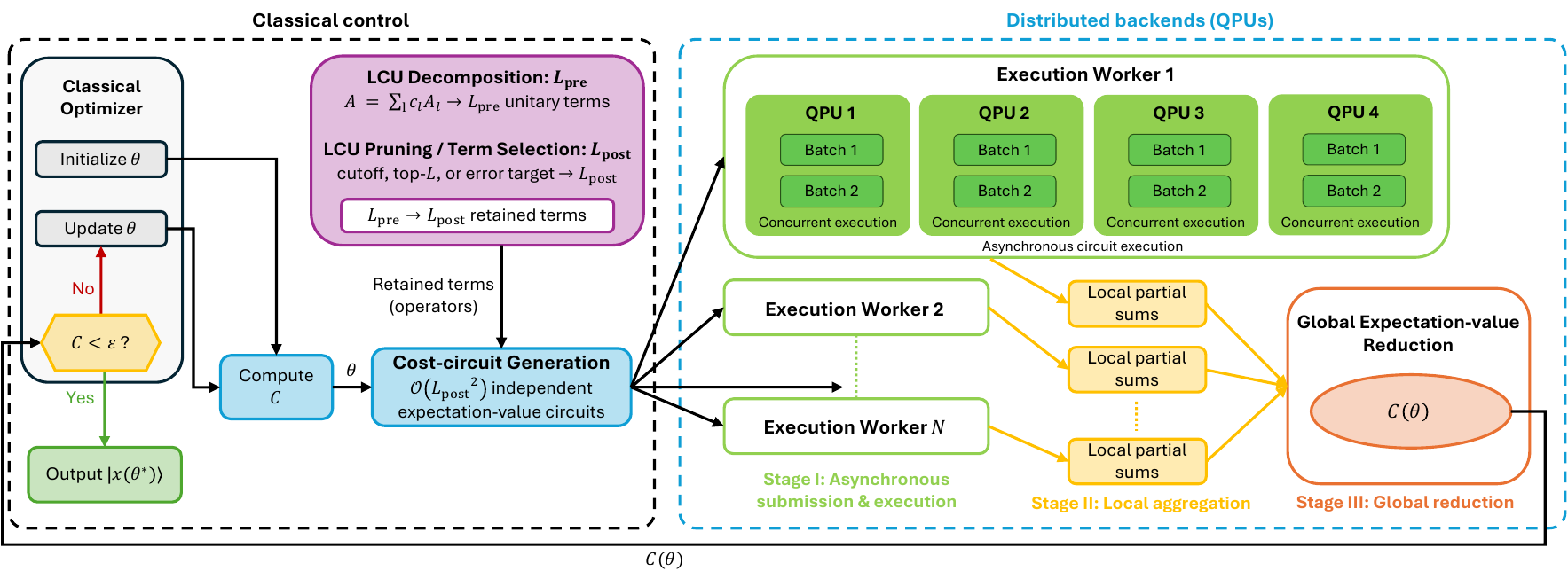}
    \caption{Overview of the optimizer- and backend-agnostic D-VQLS framework. Matrix preprocessing decomposes $A$ into $L_{\mathrm{pre}}$ unitary terms and optionally prunes or selects $L_{\mathrm{post}}$ retained terms. Independently, the optimizer initializes or updates $\theta$. The retained operators and $\theta$ jointly define $\mathcal{O}(L_{\mathrm{post}}^2)$ cost circuits, which are partitioned across execution workers (Stage~I), aggregated locally (Stage~II), and reduced to $C_{L_{\mathrm{post}}}(\theta)$ (Stage~III) for the optimizer feedback loop.}
    \label{fig:distributed_schematic}
\end{figure*}

To relate pruning to solution quality, let $\tilde{A}=A+E$ denote the pruned matrix and let $x=A^{-1}b$ and $\tilde{x}=\tilde{A}^{-1}b$. Standard perturbation theory~\cite{higham2002accuracy} gives, provided $\kappa(A)\lVert E\rVert_2/\lVert A\rVert_2<1$,
\begin{equation}\label{equ:pruning_perturbation_bound}
\frac{\lVert x-\tilde{x}\rVert_2}{\lVert x\rVert_2}
\le
\frac{\kappa(A)\,\dfrac{\lVert E\rVert_2}{\lVert A\rVert_2}}
{1-\kappa(A)\,\dfrac{\lVert E\rVert_2}{\lVert A\rVert_2}}
=\mathcal{O}\!\left(\kappa(A)\frac{\lVert E\rVert_2}{\lVert A\rVert_2}\right).
\end{equation}
Thus, for a fixed pruning error, a larger condition number makes the solution more sensitive to discarded Pauli terms; well-conditioned systems can tolerate more aggressive pruning. Because this bound is conservative and depends on the right-hand side, Section~\ref{sec:pruning_results} reports the observed matrix and solution errors.

\subsection{Distributed VQLS (D-VQLS)} \label{sec:distributed}

The LCU-based cost evaluation exposes an embarrassingly parallel structure. The $\mathcal{O}(nL^2)$ expectation values in Equation~\eqref{equ:expanded_cost} are mutually independent and can therefore be evaluated concurrently. The D-VQLS framework exploits this parallelism through a distributed, asynchronous design that partitions circuit evaluations across available compute ranks, each backed by one or more quantum processing backends. In this work, we implement D-VQLS using NVIDIA CUDA-Q~\cite{kim2023cuda}. Within each rank, the multi-QPU (\texttt{mqpu}) backend exposes a set of independently addressable QPUs to which circuits are dispatched asynchronously; MPI provides the classical coordination among ranks, distributing the workload and reducing the resulting partial sums. Here, each QPU is simulated by a GPU-accelerated state-vector backend; the
same execution model extends naturally to physical quantum processors as they become available. 

The evaluation of the cost function is organized into three steps:
\begin{enumerate}[label=\roman*.]
\item \textbf{Asynchronous Submission:}  The $L^2$ ordered index pairs $(l,l')$ arising in Equation~\eqref{equ:expanded_cost} are distributed across the available ranks using a strided workload allocation. For each assigned pair, a rank submits the corresponding Hadamard-test circuits, comprising $n$ circuits estimating the numerator overlaps and one circuit estimating the denominator overlap, with separate circuits for the real and imaginary parts of each, totaling $2(n+1)$ circuits per $(l,l')$ pair as discussed in Section~\ref{sec:methodology}. Circuits are submitted asynchronously via the \texttt{cudaq.observe\_async} API and assigned to the rank's QPUs in round-robin fashion.

\item \textbf{Local Aggregation:} Each rank waits for its asynchronous
futures to resolve, combines the real and imaginary Hadamard-test outcomes into a complex expectation value for each overlap, and accumulates the weighted partial sums $E_{\text{loc}}$ and $\Psi_{\text{loc}}$ of the numerator and denominator contributions, with weights $c_l^* c_{l'}$.

\item \textbf{Global Reduction:} A global synchronization step via \texttt{MPI\_Allreduce} sums $E_{\text{loc}}$ and $\Psi_{\text{loc}}$ across all ranks, yielding the totals $E$ and $\Psi$, after which every rank evaluates the cost in the closed form $C_L = \tfrac{1}{2} - \tfrac{1} {2}\,\mathrm{Re}\!\left[E/(n\Psi)\right]$. Only two complex scalars are communicated per cost evaluation, so the inter-rank communication volume is independent of both the qubit count $n$ and the number of retained LCU terms $L$.
\end{enumerate}

\begin{figure*}[t]
    \centering
    \begin{minipage}[t]{0.32\textwidth}
        \centering
        \includegraphics[width=\linewidth]{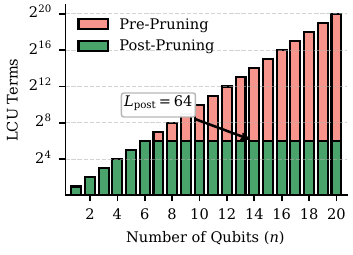}
        \vspace{2pt}
        {\footnotesize (a) LCU terms before and after pruning.\par}
    \end{minipage}\hfill
    \begin{minipage}[t]{0.32\textwidth}
        \centering
        \includegraphics[width=\linewidth]{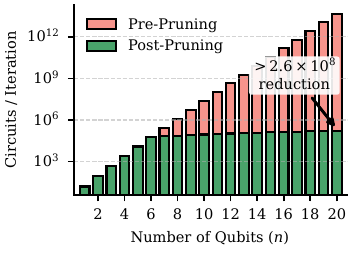}
        \vspace{2pt}
        {\footnotesize (b) Circuits per optimizer iteration.\par}
    \end{minipage}\hfill
    \begin{minipage}[t]{0.32\textwidth}
        \centering
        \includegraphics[width=\linewidth]{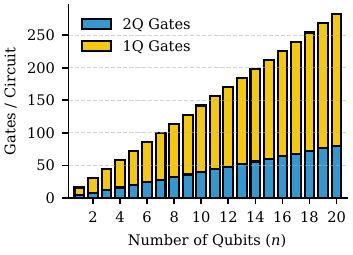}
        \vspace{2pt}
        {\footnotesize (c) One- and two-qubit gates per circuit.\par}
    \end{minipage}
    \caption{Raw circuit-resource scaling for the tridiagonal Toeplitz benchmark under the coefficient-amplitude pruning $\tau=0.01$. The full decomposition contains $L_{\rm pre}=2^n$ terms, whereas pruning retains $L_{\rm post}=\min(2^n,64)$; consequently, the circuit count $2(n+1)L^2$ changes from exponential to linear growth after $n=6$. Here, $\tau$ limits each discarded coefficient and is not a Frobenius-error tolerance. Gate counts exclude the input-dependent state-preparation block $U_b$. These are circuit-level estimates before error-correction overheads.}
    \label{fig:combined_resources}
\end{figure*}

The three steps above are repeated for every optimizer iteration. Figure~\ref{fig:distributed_schematic} illustrates the end-to-end D-VQLS workflow, from the pruned LCU terms $\{c_l, A_l\}$ and current ansatz parameters $\theta$ through distributed evaluation to the cost $C_L$ returned to the classical optimizer.

Algorithm~\ref{alg:vqls} presents the complete D-VQLS procedure. The LCU decomposition and pruning are performed once as classical preprocessing, and the resulting terms $\{c_l, A_l\}$ are broadcast to all ranks during initialization. Each Hadamard-test circuit comprises a $d$-layer hardware-efficient ansatz $V(\theta)$ on the $n$ system qubits, two controlled Pauli-string operators implementing $A_l^\dagger$ and $A_{l'}$, the controlled $Z_j$ operation, and the ancilla preparation and measurement. Each ansatz layer contributes $3n$ single-qubit rotations and $n-1$ CZ gates, and each controlled Pauli string requires at most $n$ two-qubit gates, so each circuit uses $3dn + 2$ single-qubit and at most $d(n-1) + 2n + 1$ two-qubit gates on $n+1$ qubits, excluding the state-preparation blocks $U_b$ (discussed in Section~\ref{sec:resource}).

\begin{algorithm}[t]
\caption{The D-VQLS Framework}
\label{alg:vqls}
\begin{algorithmic}
\State \textbf{Input:} Matrix $A \in \mathbb{R}^{2^n \times 2^n}$, vector $\mathbf{b}$, ansatz depth $d$, optimizer tolerance $\varepsilon$, pruning threshold $\tau$
\State \textbf{Output:} Approximate solution state $|x(\theta^*)\rangle = V(\theta^*)|0\rangle$

\State \textbf{Step 1:} Decompose $A = \sum_{l=1}^{L} c_l A_l$ via FWHT-based Pauli decomposition
\State \textbf{Step 2:} Scan once and retain $|c_l|>\tau$
\State \textbf{Step 3:} Initialize $\theta$ randomly; distribute $\{c_l, A_l\}$ to all ranks

\While{$C(\theta) > \varepsilon$}

    \State \textbf{Step 4a --- Asynchronous Circuit Dispatch:}
    \State Distribute $(l,l')$ pairs across ranks; each rank submits:
    \State \hspace{1.5em} Numerator circuits: $\langle x|A_l^\dagger P_j A_{l'}|x\rangle$, $j=1,\dots,n$
    \State \hspace{1.5em} Denominator circuits: $\langle x|A_l^\dagger A_{l'}|x\rangle$

    \State \textbf{Step 4b --- Local Aggregation:}
    \State $E_{\text{loc}} \leftarrow \sum_{(l,l')\in\text{rank}} c_l^*\,c_{l'}\,(\text{Re}+i\,\text{Im})_{\text{num}}$
    \State $\Psi_{\text{loc}} \leftarrow \sum_{(l,l')\in\text{rank}} c_l^*\,c_{l'}\,(\text{Re}+i\,\text{Im})_{\text{den}}$

    \State \textbf{Step 4c --- Global Reduction:}
    \State $(E, \Psi) \leftarrow \texttt{Allreduce}(E_{\text{loc}}, \Psi_{\text{loc}})$
    \State $C(\theta) \leftarrow \frac{1}{2} - \frac{1}{2}\,\left(\frac{E}{n \cdot \Psi}\right)$
    \State $\theta \leftarrow \text{Optimizer}(\theta, C(\theta))$

\EndWhile

\State \textbf{Step 5:} 
$\theta^* \leftarrow \theta$
\Return $|x(\theta^*)\rangle = V(\theta^*)|0\rangle$
\end{algorithmic}
\end{algorithm}

\subsection{Applications} \label{sec:applications}

\subsubsection{Tridiagonal Toeplitz Linear System}

Tridiagonal Toeplitz matrices arise frequently in the finite-difference discretization of second-order ordinary differential equations and have broad applications across scientific computing~\cite{noschese2013tridiagonal}. The matrix $A \in \mathbb{R}^{N \times N}$ takes the form:
\begin{equation}
A = \begin{pmatrix}
a & b & 0 & \cdots & 0 \\
c & a & b & \ddots & \vdots \\
0 & c & a & \ddots & 0 \\
\vdots & \ddots & \ddots & \ddots & b \\
0 & \cdots & 0 & c & a
\end{pmatrix}
\end{equation}
where $a$, $b$, and $c$ are scalar constants on the main diagonal, super-diagonal, and sub-diagonal, respectively.

Although the FWHT procedure is output-efficient, avoiding dense matrix factorization, an arbitrary matrix can still have many nonzero Pauli coefficients. For the experiments, we normalize the largest Toeplitz coefficient to one and discard each coefficient satisfying $|c_l|\le\tau=0.01$. The value $\tau=0.01$ is a coefficient-amplitude pruning, not a claim that the pruned matrix has $1\%$ error: many discarded coefficients can contribute to the cumulative matrix error. This is an empirical operating point rather than a universal constant. For the tridiagonal Toeplitz family, this threshold retains $2^n$ Pauli terms for $n\leq6$ and saturates at 64 terms for $n>6$. Section~\ref{sec:pruning_results} measures the resulting cumulative matrix and solution errors, and Section~\ref{sec:scalability} reports how the retained-term count and distributed workload change with $\tau$.

\subsubsection{Hele--Shaw Flow}

As a physically motivated benchmark, we consider the pressure-driven Hele--Shaw flow, a canonical two-dimensional creeping flow confined between two closely spaced parallel plates. Together with the tridiagonal Toeplitz system, this benchmark connects the validation cases to linear-solve kernels used in scientific workflows: the Toeplitz matrix is the canonical finite-difference Poisson stencil, while the Hele--Shaw pressure solve represents the pressure-correction step that appears in incompressible-flow time integration. The governing equations reduce to a set of Poisson equations for the velocity ($\boldsymbol{u}$) and pressure ($p$) fields, given by
\begin{align}
    \nabla^2 \boldsymbol{u} = \nabla p\\
    \nabla^2 p = 0
\end{align}
subject to appropriate Dirichlet and Neumann boundary conditions. Discretizing this equation on a uniform grid via second-order finite differences yields a sparse, symmetric positive-definite linear system $A\mathbf{x} = \mathbf{b}$, where $\mathbf{x}$ represents the unknown pressure values at interior grid points. This problem has been previously studied in the context of quantum linear solvers~\cite{meena2024solving, meena2025assessing, lu2020quantum, lapworth2022hybrid} and serves as a representative benchmark for evaluating VQLS on physically relevant linear systems with non-trivial structure.

\subsection{Resource Estimation} \label{sec:resource}

Figure~\ref{fig:combined_resources} summarizes the raw circuit-resource estimates for the D-VQLS Hadamard-test circuit applied to the tridiagonal Toeplitz benchmark for $n=1$ to $20$ qubits. These are algorithmic circuit requirements rather than fault-tolerant estimates and therefore exclude quantum error correction, logical-qubit layout, routing, and physical-qubit overheads. Each circuit requires $n+1$ qubits ($n$ system qubits plus one Hadamard-test ancilla). With the coefficient pruning $\tau=0.01$ used in this work, the retained term count is $L_{\rm post}=\min(2^n,64)$, giving $2(n+1)L_{\rm post}^2$ circuit submissions per optimizer iteration. For the 10-qubit case used in the scaling experiments, this corresponds to 64 retained LCU terms and 90,112 circuits per optimizer iteration.

The gate-count plot reports the D-VQLS Hadamard-test circuit body only. This body includes the hardware-efficient ansatz $V(\theta)$, the controlled Pauli-string operator $(A_l^\dagger A_{l'})$, and the Hadamard-test ancilla operations. State preparation remains necessary to realize $|b\rangle = U_b|0\rangle$ when the cost-function circuit calls for $U_b$, but it is treated as a separate input-loading block because its implementation depends on the vector $\mathbf{b}$ and the chosen encoding strategy rather than on the Toeplitz LCU pruning studied here. For an arbitrary unstructured dense matrix, the full Pauli basis can still require $L = 4^n$ LCU terms in the worst case, giving $2(n+1) \cdot 16^n$ circuit submissions per optimizer iteration; the Toeplitz curves in Figure~\ref{fig:combined_resources} therefore depend on Pauli-coefficient compressibility rather than on sparsity alone.


\section{Results} \label{sec:result}

\begin{figure}[t]
  \centering
  \includegraphics[width=\columnwidth,trim=70 25 30 25,clip]{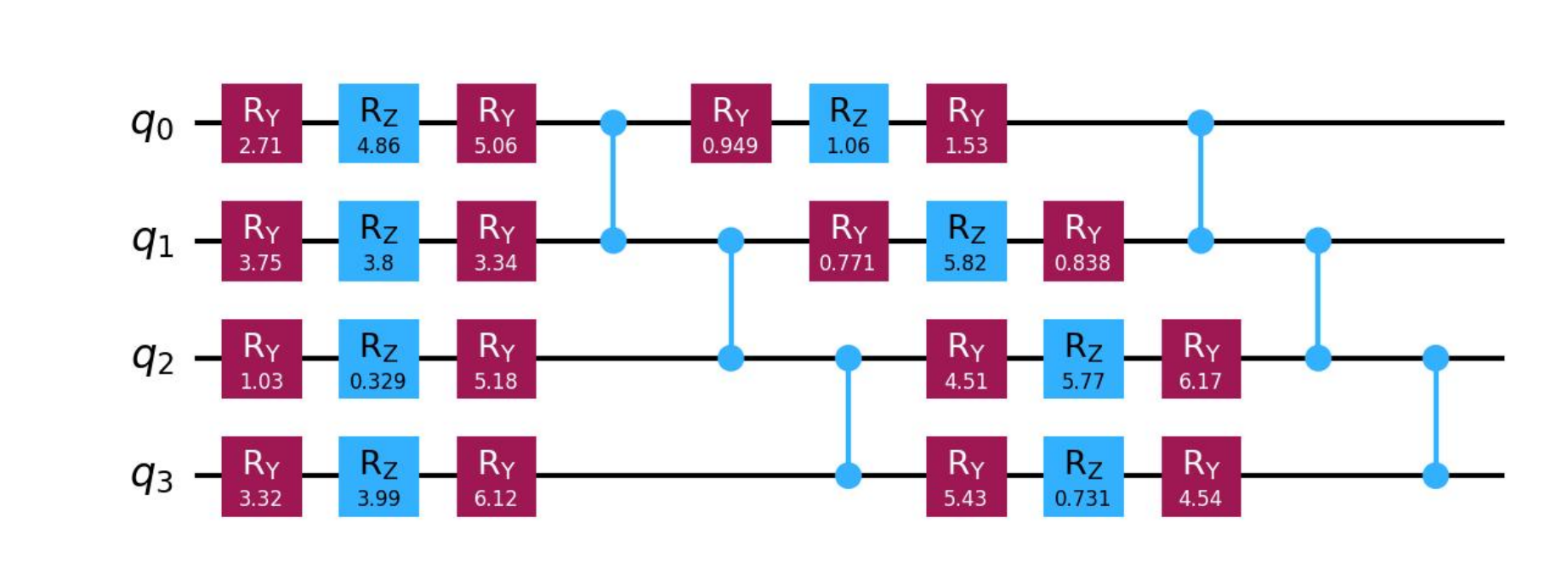}
  \caption{Hardware-efficient quantum ansatz used in all experiments \cite{lu2026memory}. Trainable parameters are encoded in the $R_y$ and $R_z$ single-qubit rotation gates, while CNOT gates provide inter-qubit entanglement.}
  \label{fig:ansatz}
\end{figure}

All experiments were conducted on the NERSC Perlmutter supercomputer. Each compute node has one 64-core AMD EPYC 7763 CPU and four NVIDIA A100-SXM4-40GB GPUs. The Anaconda environment includes Python 3.12, CUDA-Q 0.12, and SciPy 1.16. We use the limited-memory Broyden--Fletcher--Goldfarb--Shanno algorithm with bounds (L-BFGS-B) for optimization. CUDA-Q's multi-QPU simulation backend, \texttt{mqpu}, enables embarrassingly parallel VQLS circuit evaluation across distributed GPU resources. Figure~\ref{fig:ansatz} shows the hardware-efficient ansatz used in all experiments \cite{lu2026memory}.

\subsection{Validation} \label{sec:validation}

To verify the correctness of the D-VQLS framework, we solved two benchmark problems: (i) a tridiagonal Toeplitz linear system and (ii) a canonical Hele--Shaw flow on a $4\times 4$ interior grid (4 qubits). For each problem, we compared the quantum solution $\ket{x(\theta^*)}$ recovered from the optimized ansatz against the classical solution obtained via direct matrix inversion, computing the state fidelity $F(|\psi\rangle, |\phi\rangle) = |\langle \psi | \phi \rangle|^2$.

\begin{figure}[t]
  \centering
  \includegraphics[width=\columnwidth]{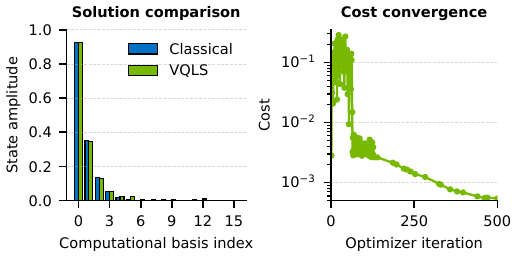}
  \caption{Solution comparison and optimizer learning curve for the 4-qubit tridiagonal Toeplitz system. Left: state-amplitude comparison between the VQLS output and the classical solution. Right: logarithmic-scale cost convergence; the optimizer reaches $10^{-3}$ cost within $\sim$250 optimizer iterations.}
  \label{fig:tridiag_validation}
\end{figure}

\begin{figure*}[t]
  \centering
  \begin{minipage}[t]{0.485\textwidth}
    \centering
    \includegraphics[width=\linewidth]{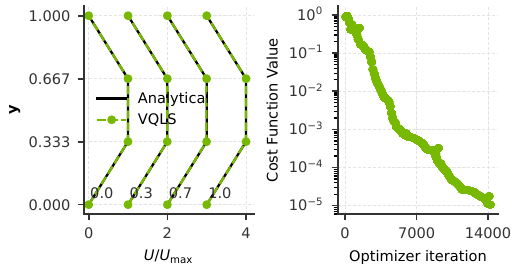}
    \caption{Hele--Shaw velocity field validation. Left: Hele--Shaw velocity plot with $4\times 4$ grid points; VQLS results agree with the analytical solution with over $99.99\%$ fidelity. Right: logarithmic-scale cost convergence; the cost drops sharply within 3,000--4,000 optimizer iterations and plateaus near zero.}
    \label{fig:heleshaw_velocity}
  \end{minipage}
  \hfill
  \begin{minipage}[t]{0.485\textwidth}
    \centering
    \includegraphics[width=\linewidth]{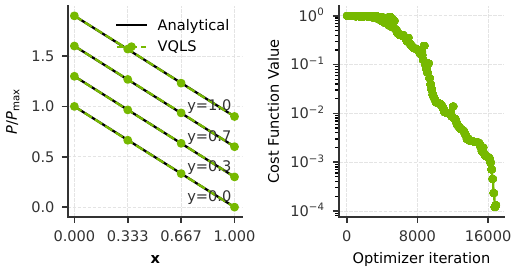}
    \caption{Hele--Shaw pressure field validation. Left: Hele--Shaw pressure profile with $4\times 4$ grid points; VQLS results reproduce the analytical linear pressure distribution. Right: logarithmic-scale cost convergence; the pressure field requires $\sim$9,000--10,000 optimizer iterations.}
    \label{fig:heleshaw_pressure}
  \end{minipage}
\end{figure*}

All benchmark problems reach near-zero cost and solution fidelity above 0.9999 relative to the classical solution, confirming the D-VQLS implementation. Figure~\ref{fig:tridiag_validation} shows that the tridiagonal Toeplitz optimizer converges within approximately 100 optimizer iterations. Figures~\ref{fig:heleshaw_velocity} and~\ref{fig:heleshaw_pressure} present the Hele--Shaw velocity and pressure results. The velocity field converges within 3,000--4,000 optimizer iterations. In the pressure-field run shown in Figure~\ref{fig:heleshaw_pressure}, the optimizer requires approximately 9,000--10,000 optimizer iterations; this is an empirical benchmark measurement rather than a literature estimate. The slower pressure convergence is consistent with the larger condition number of the Hele--Shaw pressure Laplacian, which makes the variational landscape harder to optimize~\cite{anthropicClaudeOpus48,openaiCodex2026}.

\begin{figure}[t]
  \centering
  \includegraphics[width=\columnwidth]{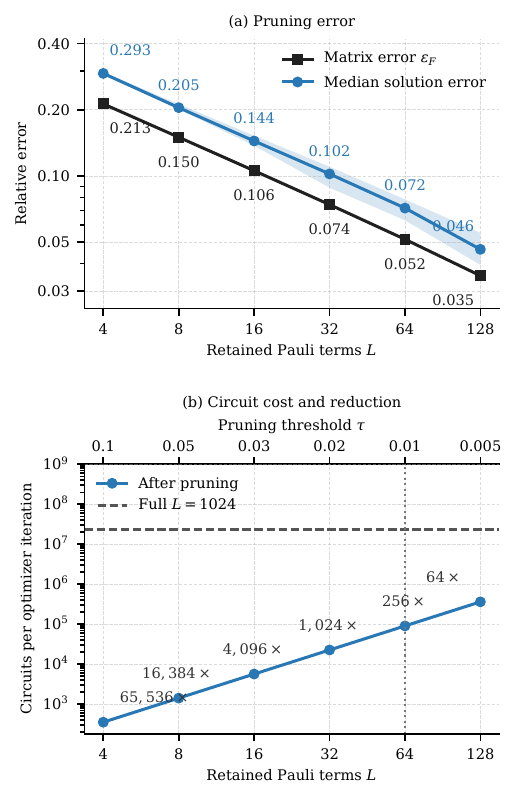}
  \caption{For the 10-qubit Toeplitz system, pruning to the $L=64$ largest-magnitude Pauli terms at $\tau=0.01$ reduces circuits per optimizer iteration by $256\times$ (23 million to 90,112) with acceptable matrix error and solution error.}
  \label{fig:pruning_analysis}
\end{figure}
\subsection{Pauli Pruning Accuracy}\label{sec:pruning_results}

To isolate Pauli-pruning error from ansatz and optimizer errors, we reconstruct each pruned matrix $\tilde{A}_L$ and solve the resulting system classically. The test (results are shown in Figure~\ref{fig:pruning_analysis}) uses a 10-qubit ($N=1024$) tridiagonal Toeplitz matrix, with diagonal entries 1 and off-diagonal entries $-1/3$. The coefficient-amplitude pruning thresholds $\tau=\{0.1,0.05,0.03,0.02,0.01,0.005\}$ retain $L=\{4,8,16,32,64,128\}$ magnitude-ranked Pauli terms. For each pruning level, we solve the full and pruned systems for 64 normalized Gaussian right-hand sides $b_j$, generated with a fixed seed. For right-hand side $b_j$, we compute $e_j(L)=\lVert \tilde{A}_L^{-1}b_j-A^{-1}b_j\rVert_2/\lVert A^{-1}b_j\rVert_2$ and report $\operatorname{median}_{j=1}^{64}e_j(L)$. The shaded interquartile band in Figure~\ref{fig:pruning_analysis}(a) spans the 25th--75th percentiles, showing the distribution across right-hand sides without emphasizing a single favorable vector.

Figure~\ref{fig:pruning_analysis}(a) shows that the matrix error decreases from $\epsilon_F=21.28\%$ at $L=4$ to $3.53\%$ at $L=128$, while the median relative solution error decreases from $29.28\%$ to $4.64\%$. Figure~\ref{fig:pruning_analysis}(b) translates this accuracy curve into execution cost. The retained circuit count grows from 352 to 360,448 circuits per optimizer iteration, but remains $65{,}536\times$ to $64\times$ smaller than the 23,068,672-circuit full decomposition. At the paper's $\tau=0.01$, $L=64$ operating point, the matrix error is $5.16\%$, the median solution error is $7.15\%$, and the circuit reduction is $256\times$.

\subsection{Scalability} \label{sec:scalability}

We evaluate the strong and weak scaling of D-VQLS using the same 10-qubit ($N=1024$) tridiagonal Toeplitz system analyzed in Section~\ref{sec:pruning_results}.

In the strong-scaling study, the total work per optimizer iteration is fixed: the 10-qubit Toeplitz problem uses 64 pruned LCU terms and therefore 90,112 Hadamard-test circuits per optimizer iteration, while the number of GPUs is increased to measure time-to-solution reduction. In the weak-scaling study, the workload per GPU is kept approximately constant at the 3,755-circuits/GPU baseline, while the coefficient-amplitude pruning threshold, retained LCU terms, total circuit count, and GPU allocation grow together. This distinction is important because strong scaling exposes the communication and synchronization floor, whereas weak scaling tests whether the framework can absorb larger circuit batches without increasing per-GPU time.

\subsubsection{Strong Scaling}

For the strong scaling study, we fix the problem size for the $1024\times 1024$ matrix at 64 pruned LCU terms, resulting in 90,112 circuits per optimizer iteration. We vary the number of GPUs from 4 (1 node) to 32 (8 nodes), using 16 MPI tasks per node (the optimal configuration identified in Section~\ref{sec:mpi_profiling}).

\begin{figure}[t]
 \centering
 \includegraphics[width=0.90\columnwidth]{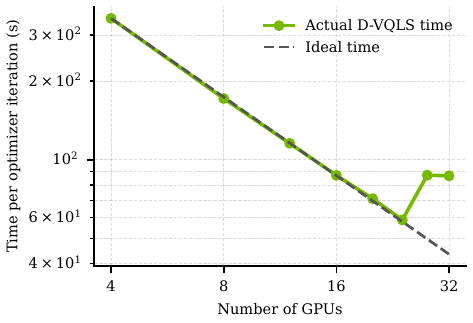}
 \caption{Strong scaling of the D-VQLS framework for a 10-qubit tridiagonal Toeplitz system (90,112 circuits per optimizer iteration) from 4 to 32 GPUs on Perlmutter, shown on log--log axes. Near-ideal scaling is observed up to 24 GPUs.}
 \label{fig:strong_VQLS}
\end{figure}

Figure~\ref{fig:strong_VQLS} presents the average wall-clock time per D-VQLS iteration as a function of GPU count. The framework exhibits near-ideal strong scaling up to 24 GPUs, at which point each GPU processes approximately 3,755 circuits per optimizer iteration. Beyond this threshold, a distinct performance inflection occurs: execution times increase when scaling to 28 and 32 GPUs. This degradation is attributable to the communication overhead of inter-node \texttt{MPI\_Allreduce} operations, which begins to dominate the reduced per-GPU compute workload. When fewer than approximately 2,800 circuits remain per GPU, the collective synchronization cost across the multi-node interconnect exceeds the time savings from further parallelization.

\subsubsection{Weak Scaling}

For the weak scaling analysis, we increase both the problem size (number of LCU terms, controlled by the coefficient-amplitude pruning threshold) and the allocated GPU resources proportionally, maintaining a constant per-GPU workload. The pruning threshold ranges from $\tau=0.1$ (4 terms) to $\tau=0.005$ (128 terms), so the total number of circuits per optimizer iteration spans three orders of magnitude from 352 to 360,448. We use the 6-node (24 GPU), 64-term configuration as the baseline, where each GPU processes 3,755 circuits.
For configurations that cannot be perfectly mapped to integer node counts at the baseline per-GPU workload, we compute the normalized time as $t_{\text{norm}} = t_{\text{actual}} \cdot N_{\text{actual}} / N_{\text{ideal}}$, where $N_{\text{actual}}$ and $N_{\text{ideal}}$ are the actual and ideal GPU allocations, respectively. This normalization accounts for the discrete allocation constraints of the computing system.

\begin{figure}[t]
  \centering
  \includegraphics[width=0.90\columnwidth]{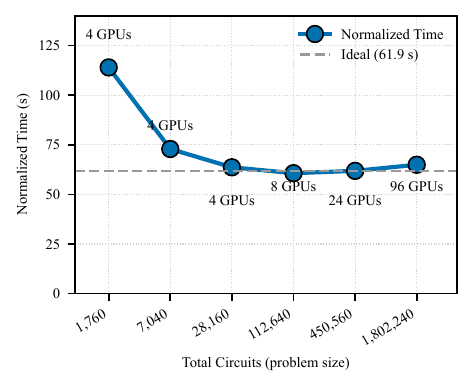}
  \caption{Weak scaling performance of the D-VQLS framework. Normalized execution time per optimizer iteration is plotted against the number of LCU terms (proportional to GPU count). The dashed line indicates ideal weak scaling. For under-saturated GPU configurations (4 and 8 LCU terms), fixed overheads inflate the normalized time; at saturation (16--128 terms), the results track the ideal line closely, achieving 95.3\% efficiency at 96 GPUs.}
  \label{fig:weak_scaling}
\end{figure}

Figure~\ref{fig:weak_scaling} presents the weak scaling results. At small problem sizes (4 and 8 LCU terms), the normalized execution times significantly exceed the ideal scaling line. This behavior is an expected artifact of hardware under-utilization: with only 88 or 352 circuits per GPU, the fixed overheads of CUDA context initialization, kernel launch latencies, and operating system scheduling dominate the execution time. As the problem size grows, the per-GPU workload saturates the available compute capacity and the normalized time approaches the ideal weak-scaling trend.

For the three configurations with saturated GPU workloads (16, 32, and 64 LCU terms, corresponding to 5,632 to 90,112 total circuits), the normalized execution times align closely with the ideal weak scaling line, demonstrating that the distributed framework introduces minimal overhead as both the problem size and compute resources grow in tandem. At the largest configuration with 128 LCU terms---distributing 360,448 circuits across 96 GPUs---the measured time of 64.9~s yields a scaling efficiency of 95.3\% relative to the baseline. These results confirm that the communication and scheduling overhead of the distributed framework remains negligible at scale.

\subsection{MPI Task Density and Load Balancing}\label{sec:mpi_profiling}

We profile \texttt{--ntasks-per-node} (ntpn $\in \{4,8,16,32\}$) on one Perlmutter node (four NVIDIA A100 GPUs). CUDA MPS maps $\text{ntpn}/4$ MPI ranks to each GPU, and NVIDIA Nsight Systems provides the CUDA API and kernel metrics.

\subsubsection{Wall-Clock Performance}

\begin{figure}[t]
\centering
\includegraphics[width=\linewidth]{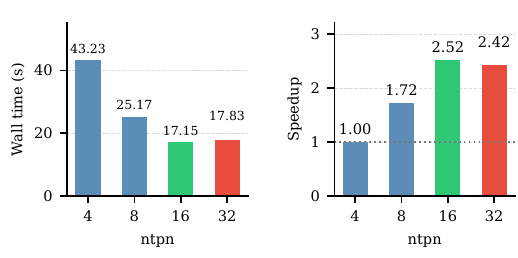}
\caption{Average wall-clock time per optimizer iteration (left) and speedup relative to ntpn$=$4 (right). ntpn$=$16 achieves the minimum wall time (17.15~s); ntpn$=$32 regresses due to MPS contention.}
\label{fig:ntpn_walltime}
\end{figure}

Figure~\ref{fig:ntpn_walltime} shows that ntpn$=$16 minimizes the wall time at 17.15~s, a $2.52\times$ speedup over ntpn$=$4. Increasing to ntpn$=$32 (8 MPS clients per GPU) regresses to 17.83~s, so ntpn$=$16 is the throughput optimum for this workload.

\subsubsection{GPU Kernel Analysis}

\begin{figure}[t]
\centering
\includegraphics[width=0.9\linewidth]{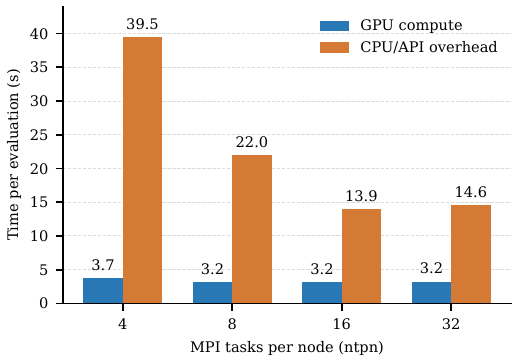}
\caption{Decomposition of wall time per optimizer iteration into GPU compute and CPU/API overhead. GPU compute time is constant ($\approx$3.1--3.7~s); the $2.5\times$ speedup from ntpn$=$4 to ntpn$=$16 is entirely due to reduced CPU-side dispatch overhead.}
\label{fig:compute_vs_overhead}
\end{figure}

The average \texttt{constMatApplKernel} latency remains 8.18--8.22~$\mu$s across configurations. Figure~\ref{fig:compute_vs_overhead} likewise shows nearly constant GPU compute time (3.1--3.7~s), whereas CPU/API overhead falls from 39.5~s at ntpn$=$4 to 13.9~s at ntpn$=$16 and rises to 14.6~s at ntpn$=$32. The speedup therefore comes from more efficient dispatch rather than faster kernels.

\subsubsection{CUDA API Overhead and MPS Contention}

\begin{figure}[t]
\centering
\includegraphics[width=\linewidth]{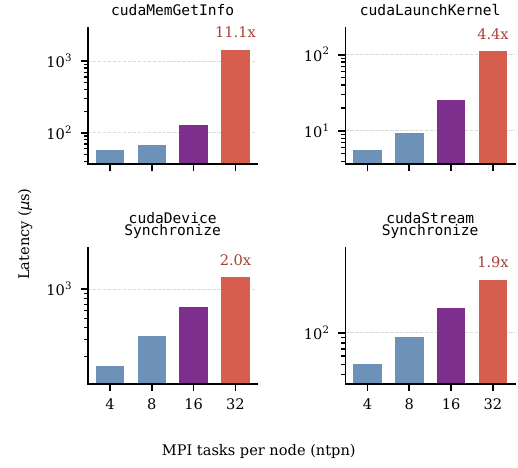}
\caption{CUDA API call latency (log scale) across configurations. The transition from ntpn$=$16 to ntpn$=$32 crosses an MPS contention threshold: \texttt{cudaMemGetInfo} latency increases by $11\times$, and \texttt{cudaLaunchKernel} latency increases by $4.4\times$.}
\label{fig:cuda_api_overhead}
\end{figure}

Figure~\ref{fig:cuda_api_overhead} identifies the ntpn$=$32 MPS contention threshold. MPS lets multiple MPI ranks share a GPU and submit work concurrently, which initially overlaps host dispatch and improves utilization. Once the client count exceeds the useful concurrency, however, ranks compete for finite driver, command-queue, and execution resources; API requests queue and synchronization waits lengthen even when the device kernels themselves do not slow down. From ntpn$=$16 to 32, average \texttt{cudaMemGetInfo} latency rises from 130 to 1,439~$\mu$s ($11.1\times$), \texttt{cudaLaunchKernel} rises from 25.5 to 111.8~$\mu$s ($4.4\times$), and synchronization calls approximately double. Together with the nearly constant kernel latency, these increases show that host-side queuing at 8 clients per GPU erases the benefit of additional ranks.

Together, the wall-time, kernel, and API profiles establish ntpn$=$16 (4 MPS clients per GPU) as the last efficient concurrency point before MPS contention dominates.

\section{Conclusion and Discussion} \label{sec:conclusion}

This work presented D-VQLS, the first scalable, distributed Variational Quantum Linear Solver that addresses the algorithm's primary practical bottleneck: the circuit-execution overhead of the Linear Combination of Unitaries (LCU) decomposition. The FWHT-based Pauli decomposition provides an additional acceleration mechanism for the D-VQLS framework. We built an asynchronous, MPI-based framework on NVIDIA CUDA-Q that distributes circuit evaluations across multi-node, multi-GPU HPC resources and exploits the embarrassingly parallel structure of the LCU-based cost function.

FWHT-based Pauli decomposition and pruning provide the preprocessing stage of the end-to-end acceleration. For the 10-qubit Toeplitz benchmark, retaining $L=64$ reduces the circuit count per optimizer iteration from 23{,}068{,}672 to 90{,}112 ($256\times$); D-VQLS then distributes the remaining workload. Because the decomposition is reused across optimizer iterations and can be amortized across right-hand sides with the same $A$, preprocessing and distributed execution provide complementary algorithmic and systems-level gains~\cite{anthropicClaudeOpus48,openaiCodex2026}.

The framework achieves near-ideal strong scaling up to 24 GPUs on a 10-qubit, 90{,}112-circuit workload and 95.3\% weak-scaling efficiency at 96 GPUs (360{,}448 circuits per optimizer iteration), confirming minimal overhead as problem size and resources grow together. Nsight Systems profiling further identifies an optimal intra-node density of 16 MPI tasks per node (4 MPS clients per A100 GPU), a $2.52\times$ speedup over the single-client baseline. These results establish that VQLS with general LCU-encoded matrices scales efficiently on current HPC systems, making the workload amenable to distributed execution.

All results in this work are based on noiseless GPU state-vector simulation, so hardware noise and communication effects remain outside the present scope. The asynchronous execution model should carry over to any backend that can run independent circuits concurrently. The effectiveness of Pauli-coefficient pruning is problem dependent: structured matrices with low condition numbers can tolerate more aggressive pruning, whereas poorly conditioned systems are more sensitive to the resulting perturbation. An immediate next step is to evaluate these effects on QPU hardware and across a broader set of linear systems, including the role of preconditioning and alternative matrix decompositions.

\section*{Acknowledgments}
This research used resources of the National Energy Research Scientific Computing Center (NERSC), a U.S. Department of Energy Office of Science User Facility at Lawrence Berkeley National Laboratory under Contract No. DE-AC02-05CH11231, through awards ASCR-ERCAP0034931 and ASCR-ERCAP0036194; and the Oak Ridge Leadership Computing Facility at Oak Ridge National Laboratory, which is supported by the Advanced Scientific Computing Research programs in the Office of Science of the U.S. Department of Energy under Contract No. DE-AC05-00OR22725.
Portions of this manuscript were prepared with the assistance of large language model (LLM) tools, including Anthropic's Claude code and OpenAI Codex, which were used to support drafting, editing, grammar refinement, and stylistic improvements of the text. The authors also used these tools to assist with formatting LaTeX source, polishing figure captions, and organizing prose for clarity and conciseness. All AI-generated content was reviewed, verified, and edited by the authors, who take full responsibility for the technical accuracy, scientific claims, experimental results, data analysis, and conclusions presented in this work. No AI tools were used to generate or fabricate experimental data, numerical results, figures, or citations; all references were independently verified by the authors.
\balance
\bibliographystyle{IEEEtran}
\bibliography{references}

@article{lu2026memory,
  title={Memory-, Circuit-, and Ansatz-Efficient VQLS for CFD on Hybrid Quantum-HPC Systems},
  author={Lu, Chao and Gopalakrishnan Meena, Muralikrishnan and Perez, Eduardo Antonio Coello and Gottiparthi, Kalyana Chakravarthi and Kim, Seongmin},
  journal={arXiv preprint arXiv:2608.09661},
  year={2026}
}

@inproceedings{kim2023cuda,
  title={Cuda quantum: The platform for integrated quantum-classical computing},
  author={Kim, Jin-Sung and McCaskey, Alex and Heim, Bettina and Modani, Manish and Stanwyck, Sam and Costa, Timothy},
  booktitle={2023 60th ACM/IEEE Design Automation Conference (DAC)},
  pages={1--4},
  year={2023},
  organization={IEEE},
  url={https://doi.org/10.1109/DAC56929.2023.10247886}
}

@inproceedings{meena2025assessing,
  title={Assessing VQLS for Fluid Dynamics on a Hybrid Quantum-HPC Stack},
  author={Gopalakrishnan Meena, Muralikrishnan and Lu, Chao and P{\'e}rez, Eduardo Antonio Coello and Shehata, Amir and Kim, Seongmin and Gottiparthi, Kalyana Chakravarthi and Suh, In-Saeng},
  booktitle={2025 IEEE International Conference on Quantum Computing and Engineering (QCE)},
  volume={2},
  pages={484--485},
  year={2025},
  organization={IEEE}
}

@article{bosco2024demonstration,
  title={Demonstration of Scalability and Accuracy of Variational Quantum Linear Solver for Computational Fluid Dynamics},
  author={Bosco, Ferdin Sagai Don and Lineswala, Rut and Chopra, Abhishek and others},
  journal={arXiv preprint arXiv:2409.03241},
  year={2024},
  url={https://arxiv.org/abs/2409.03241}
}

@article{ye2024hybrid,
  title={A hybrid quantum-classical framework for computational fluid dynamics},
  author={Ye, Chuang-Chao and An, Ning-Bo and Ma, Teng-Yang and Dou, Meng-Han and Bai, Wen and Sun, De-Jun and Chen, Zhao-Yun and Guo, Guo-Ping},
  journal={Physics of Fluids},
  volume={36},
  number={12},
  year={2024},
  publisher={AIP Publishing},
  url={https://doi.org/10.1063/5.0238193}
}

@article{demirdjian2022variational,
  title={Variational quantum solutions to the advection--diffusion equation for applications in fluid dynamics},
  author={Demirdjian, Reuben and Gunlycke, Daniel and Reynolds, Carolyn A and Doyle, James D and Tafur, Sergio},
  journal={Quantum Information Processing},
  volume={21},
  number={9},
  pages={322},
  year={2022},
  publisher={Springer},
  url={https://doi.org/10.1007/s11128-022-03667-7}
}

@article{golden2022quantum,
  title={{Quantum computing and preconditioners for hydrological linear systems}},
  author={Golden, John and O’Malley, Daniel and Viswanathan, Hari},
  journal={Scientific Reports},
  volume={12},
  number={1},
  pages={22285},
  year={2022},
  publisher={Nature Publishing Group UK London},
  url={https://doi.org/10.1038/s41598-022-25727-9}
}

@book{trefethen2022numerical,
  title={Numerical linear algebra},
  author={Trefethen, Lloyd N and Bau, David},
  year={2022},
  publisher={SIAM}
}

@book{higham2002accuracy,
  title={Accuracy and Stability of Numerical Algorithms},
  author={Higham, Nicholas J.},
  edition={2},
  year={2002},
  publisher={SIAM},
  address={Philadelphia, PA},
  url={https://doi.org/10.1137/1.9780898718027}
}

@article{bravo2023variational,
  title={Variational quantum linear solver},
  author={Bravo-Prieto, Carlos and LaRose, Ryan and Cerezo, Marco and Subasi, Yigit and Cincio, Lukasz and Coles, Patrick J},
  journal={Quantum},
  volume={7},
  pages={1188},
  year={2023},
  publisher={Verein zur F{\"o}rderung des Open Access Publizierens in den Quantenwissenschaften},
  url={https://doi.org/10.22331/q-2023-11-22-1188}
}

@article{lin2020optimal,
  title={Optimal polynomial based quantum eigenstate filtering with application to solving quantum linear systems},
  author={Lin, Lin and Tong, Yu},
  journal={Quantum},
  volume={4},
  pages={361},
  year={2020},
  publisher={Verein zur F{\"o}rderung des Open Access Publizierens in den Quantenwissenschaften},
  url={https://doi.org/10.22331/q-2020-11-11-361}
}

@article{costa2022optimal,
  title={Optimal scaling quantum linear-systems solver via discrete adiabatic theorem},
  author={Costa, Pedro CS and An, Dong and Sanders, Yuval R and Su, Yuan and Babbush, Ryan and Berry, Dominic W},
  journal={PRX quantum},
  volume={3},
  number={4},
  pages={040303},
  year={2022},
  publisher={APS},
  url={https://doi.org/10.1103/PRXQuantum.3.040303}
}

@article{an2022quantum,
  title={Quantum linear system solver based on time-optimal adiabatic quantum computing and quantum approximate optimization algorithm},
  author={An, Dong and Lin, Lin},
  journal={ACM Transactions on Quantum Computing},
  volume={3},
  number={2},
  pages={1--28},
  year={2022},
  publisher={ACM New York, NY},
  url={https://doi.org/10.1145/3498331}
}

@article{subacsi2019quantum,
  title={Quantum algorithms for systems of linear equations inspired by adiabatic quantum computing},
  author={Suba{\c{s}}{\i}, Yi{\u{g}}uit and Somma, Rolando D and Orsucci, Davide},
  journal={Physical review letters},
  volume={122},
  number={6},
  pages={060504},
  year={2019},
  publisher={APS},
  url={https://doi.org/10.1103/PhysRevLett.122.060504}
}

@article{lu2025lugo,
  title={{LuGo: an Enhanced Quantum Phase Estimation Implementation}},
  author={Lu, Chao and Gopalakrishanan Meena, Muralikrishnan and Gottiparthi, Kalyana Chakravarthi},
  journal={arXiv preprint arXiv:2503.15439},
  year={2025},
  url={https://arxiv.org/abs/2503.15439}
}

@article{preskill2018quantumNISQ,
  title={{Quantum computing in the NISQ era and beyond}},
  author={Preskill, John},
  journal={Quantum},
  volume={2},
  pages={79},
  year={2018},
  publisher={Verein zur F{\"o}rderung des Open Access Publizierens in den Quantenwissenschaften},
  url={https://doi.org/10.22331/q-2018-08-06-79}
}

@article{harrow2009quantum,
  title={Quantum algorithm for linear systems of equations},
  author={Harrow, Aram W and Hassidim, Avinatan and Lloyd, Seth},
  journal={Physical review letters},
  volume={103},
  number={15},
  pages={150502},
  year={2009},
  publisher={APS},
  url={https://doi.org/10.1103/PhysRevLett.103.150502}
}

@article{coppersmith1982asymptotic,
  title={On the asymptotic complexity of matrix multiplication},
  author={Coppersmith, Don and Winograd, Shmuel},
  journal={SIAM Journal on Computing},
  volume={11},
  number={3},
  pages={472--492},
  year={1982},
  publisher={SIAM},
  url={https://doi.org/10.1137/0211038}
}

@article{meena2024solving,
  title={{Solving the Hele--Shaw flow using the Harrow--Hassidim--Lloyd algorithm on superconducting devices: A study of efficiency and challenges}},
  author={Gopalakrishnan Meena, Muralikrishnan and Gottiparthi, Kalyana C and Lietz, Justin G and Georgiadou, Antigoni and Coello P{\'e}rez, Eduardo Antonio},
  journal={Physics of Fluids},
  volume={36},
  number={10},
  year={2024},
  publisher={AIP Publishing},
  url={https://doi.org/10.1063/5.0231929}
}

@inproceedings{lu2020quantum,
  title={{Quantum CFD simulations for heat transfer applications}},
  author={Lu, Chao and Hu, Zhao and Xie, Bei and Zhang, Ning},
  booktitle={ASME International Mechanical Engineering Congress and Exposition},
  volume={84584},
  pages={V010T10A050},
  year={2020},
  organization={American Society of Mechanical Engineers},
  url={https://doi.org/10.1115/IMECE2020-23915}
}

@article{lapworth2022hybrid,
  title={{A hybrid quantum-classical CFD methodology with benchmark HHL solutions}},
  author={Lapworth, Leigh},
  journal={arXiv preprint arXiv:2206.00419},
  year={2022},
  url={https://arxiv.org/abs/2206.00419}
}

@article{georges2025paulidecomp,
  title={Pauli decomposition via the fast Walsh-Hadamard transform},
  author={Georges, Timothy N and Berntson, Bjorn K and S{\"u}nderhauf, Christoph and Ivanov, Aleksei V},
  journal={New Journal of Physics},
  volume={27},
  number={3},
  pages={033004},
  year={2025},
  publisher={IOP Publishing},
  url={https://doi.org/10.1088/1367-2630/adb44d}
}

@inproceedings{gilyen2019quantum,
  title={Quantum singular value transformation and beyond: exponential improvements for quantum matrix arithmetics},
  author={Gily{\'e}n, Andr{\'a}s and Su, Yuan and Low, Guang Hao and Wiebe, Nathan},
  booktitle={Proceedings of the 51st annual ACM SIGACT symposium on theory of computing},
  pages={193--204},
  year={2019},
  url={https://doi.org/10.1145/3313276.3316366}
}

@article{chakraborty2018power,
  title={The power of block-encoded matrix powers: improved regression techniques via faster Hamiltonian simulation},
  author={Chakraborty, Shantanav and Gily{\'e}n, Andr{\'a}s and Jeffery, Stacey},
  journal={arXiv preprint arXiv:1804.01973},
  year={2018},
  url={https://arxiv.org/abs/1804.01973}
}

@article{camps2024explicit,
  title={Explicit quantum circuits for block encodings of certain sparse matrices},
  author={Camps, Daan and Lin, Lin and Van Beeumen, Roel and Yang, Chao},
  journal={SIAM Journal on Matrix Analysis and Applications},
  volume={45},
  number={1},
  pages={801--827},
  year={2024},
  publisher={SIAM},
  url={https://doi.org/10.1137/22M1484298}
}

@article{sunderhauf2024block,
  title={Block-encoding structured matrices for data input in quantum computing},
  author={S{\"u}nderhauf, Christoph and Campbell, Earl and Camps, Joan},
  journal={Quantum},
  volume={8},
  pages={1226},
  year={2024},
  publisher={Verein zur F{\"o}rderung des Open Access Publizierens in den Quantenwissenschaften},
  url={https://doi.org/10.22331/q-2024-01-11-1226}
}

@article{loaiza2025majorana,
  title={Majorana Tensor Decomposition: A unifying framework for decompositions of fermionic Hamiltonians to Linear Combination of Unitaries},
  author={Loaiza, Ignacio and Sankar Brahmachari, Aritra and Izmaylov, Artur F},
  journal={Quantum Science and Technology},
  volume={10},
  number={3},
  pages={035035},
  year={2025},
  publisher={IOP Publishing},
  url={https://doi.org/10.1088/2058-9565/adb427}
}

@article{hogancamp2026linear,
  title={A Linear Combination of Unitaries Decomposition for the Laplace Operator},
  author={Hogancamp, Thomas and Demirdjian, Reuben and Gunlycke, Daniel},
  journal={arXiv preprint arXiv:2601.06370},
  year={2026},
  url={https://arxiv.org/abs/2601.06370}
}

@article{noschese2013tridiagonal,
  title={Tridiagonal Toeplitz matrices: properties and novel applications},
  author={Noschese, Silvia and Pasquini, Lionello and Reichel, Lothar},
  journal={Numerical linear algebra with applications},
  volume={20},
  number={2},
  pages={302--326},
  year={2013},
  publisher={Wiley Online Library},
  url={https://doi.org/10.1002/nla.1811}
}

@misc{anthropicClaudeOpus48,
  author={{Anthropic}},
  title={Introducing Claude Opus 4.8},
  year={2026},
  month=may,
  url={https://www.anthropic.com/news/claude-opus-4-8},
  note={Accessed: 2026-08-17}
}

@misc{openaiCodex2026,
  author={{OpenAI}},
  title={Introducing the Codex App},
  year={2026},
  month=feb,
  url={https://openai.com/index/introducing-the-codex-app/},
  note={Accessed: 2026-08-17}
}


\end{document}